\documentclass[twocolumn,showpacs,preprintnumbers,amsmath,amssymb]{revtex4}

\usepackage{graphicx}
\usepackage{dcolumn}
\usepackage{bm}

\begin{document}


\title{Analytic solution of the separability criterion for 
continuous variable systems}

\author{Kazuo Fujikawa}
\affiliation{%
Institute of Quantum Science, College of Science and Technology,
Nihon University, Chiyoda-ku, Tokyo 101-8308, Japan
}%


\begin{abstract}
By using the algebraic separability criterion of R. Simon, we 
present an explicit determination of squeezing parameters for 
which  the P-representation condition  saturates the 
$Sp(2,R)\otimes Sp(2,R)$ invariant separability condition for 
continuous variable two-party Gaussian systems. 
We thus give for the first time the explicit analytic formulas of 
squeezing parameters which establish the 
equivalence of the separability condition with the 
P-representation condition. The implications of our algebraic analysis on some of the past related works are discussed.
\end{abstract}

\maketitle

\section{Introduction}

The entanglement is a basic notion in quantum mechanics, but the 
quantitative criterion of entanglement is known only for a 
simple system such as a two-spin 
system~\cite{peres, horodecki}. In view of this fact, it is remarkable that 
a proof of the  necessary and sufficient separability 
criterion for  continuous variable two-party Gaussian systems
has been given by R. Simon~\cite{ simon} on the basis of  generalized Peres-Horodecki criterion~\cite{peres, horodecki}. He also gave an 
algebraic criterion for separability ({\em i.e.},  
non-entanglement)~\cite{ simon}, though it was not used in 
his proof. This algebraic criterion is considered to be fundamental, but its explicit analysis has not been performed so 
far. See a review~\cite{mancini2} on the present status of 
the quantum separability problem of Gaussian systems.

We here present an explicit determination of squeezing parameters
for which the P-representation condition saturates the 
$Sp(2,R)\otimes Sp(2,R)$ invariant separability condition by 
explicitly solving the algebraic condition of Simon. 
We thus give for the first time the explicit formulas of 
squeezing parameters, which establish the equivalence of the 
separability condition with the P-representation condition, in 
terms of the parameters of the standard form of the covariance 
matrix (or second moments) for Gaussian systems.
These explicit analytic solutions should be useful in the quantitative theoretical and experimental analyses of entanglement such as in~\cite{furusawa}.

We also show that our analytic solutions of squeezing parameters
$r_{1}$ and $r_{2}$ do not satisfy  in general the equation $f(r^{\star}_{1})=0$ which appears in another formulation of the separability criterion for two-party Gaussian systems \cite{duan}. In this sense our scheme is quantitatively different from the scheme in \cite{duan}.
It is however shown that our exlicit analytic solutions allow us to construct a concrete proof of the separability criterion on the lines of  \cite{duan}. 

\section{Analytic solutions}

We start with the $4\times 4$ correlation matrix $V=(V_{\mu\nu})$ 
where
\begin{eqnarray}  
V_{\mu\nu}=\frac{1}{2}
\langle\Delta\hat{\xi}_{\mu}\Delta\hat{\xi}_{\nu}+
\Delta\hat{\xi}_{\nu}\Delta\hat{\xi}_{\mu}\rangle
=\frac{1}{2}
\langle\{\Delta\hat{\xi}_{\mu}, \Delta\hat{\xi}_{\nu}\}
\rangle
\end{eqnarray}
with $\Delta\hat{\xi}_{\mu}=\hat{\xi}_{\mu}
-\langle\hat{\xi}_{\mu}\rangle$ in term of the variables
$(\hat{\xi}_{\mu})=(\hat{q}_{1}, \hat{p}_{1}, \hat{q}_{2}, 
\hat{p}_{2})$ for a two-party system specified by canonical 
variables $(\hat{q}_{1},\hat{p}_{1})$ and 
$(\hat{q}_{2},\hat{p}_{2})$. We generally define 
$\langle\hat{O}\rangle=
{\rm Tr}\hat{\rho} \hat{O}$ by using the density matrix 
$\hat{\rho}$.
The correlation matrix is also written in the form 
\begin{eqnarray}
V=\left(\begin{array}{cc}
  A&C\\
  C^{T}&B\\
            \end{array}\right)
\end{eqnarray}
where $A$ and $B$ are $2\times 2$ real symmetric matrices and 
$C$ is a $2\times 2$ real matrix. 
 The standard form 
\begin{eqnarray}
V_{0}=\left(\begin{array}{cccc}
  a&0&c_{1}&0\\
  0&a&0&c_{2}\\
  c_{1}&0&b&0\\
  0&c_{2}&0&b\\            
  \end{array}\right)
\end{eqnarray}
is obtained from the general $V$ by applying the 
$Sp(2,R)\otimes Sp(2,R)$ transformations~\cite{simon}. 

The separability condition, which is derived from an
analysis of the non-negativity of the partial transposed 
density matrix $\hat{\rho}$, is written in the matrix notation~\cite{simon}
\begin{eqnarray}
V+\frac{i}{2}\left(\begin{array}{cc}
  J&0\\
  0&\pm J\\            
  \end{array}\right)\geq 0
\end{eqnarray}
with a  $2\times 2$ simplectic matrix
\begin{eqnarray}
J=\left(\begin{array}{cc}
  0&1\\
  -1&0\\
            \end{array}\right),
\end{eqnarray} 
or equivalently 
\begin{eqnarray}
&&d^{T}Ad+f^{T}Bf+2d^{T}Cf+g^{T}Ag+h^{T}Bh+2g^{T}Ch\nonumber\\
&&\geq |d^{T}Jg| + |f^{T}Jh|
\end{eqnarray}
with $f\sim h$ all standing for {\em arbitrary} real two component vectors.

When one regards the separability condition as a constraint on 
the range of
$|c_{1}|$ and $|c_{2}|$ in the standard form $V_{0}$ in (3),  
it is written as 
\begin{eqnarray}
&&4(ab-c_{1}^{2})(ab-c_{2}^{2})\geq (a^{2}+b^{2})+2|c_{1}c_{2}|
-\frac{1}{4},\nonumber\\
&&\sqrt{(2a-1)(2b-1)}\geq |c_{1}|+|c_{2}|
\end{eqnarray}
together with $a\geq 1/2$ and $b\geq 1/2$.
The first algebraic relation in (7), which was derived by 
Simon~\cite{simon}, essentially corresponds to 
\begin{eqnarray}
4{\rm det}[V_{0}+\frac{i}{2}\left(\begin{array}{cc}
  J&0\\
  0&\pm J\\            
  \end{array}\right)]\geq 0
\end{eqnarray}
and thus it is 
manifestly invariant under $Sp(2,R)\otimes Sp(2,R)$. The 
second condition in (7) is given by  the weaker conditions
derived from (6) with {\em subsidiary} constraints $g=J^{T}d$ and 
$h=\pm J^{T}f$, and it is used to exclude the 
solutions of the first inequality in (7) which allow  
$c^{2}_{1}\rightarrow\infty$ and 
$c^{2}_{2}\rightarrow\infty$ for fixed $a$ and $b$. The condition $a\geq \frac{1}{2}$, for example, is given by setting $g=J^{T}d$ and $h=f=0$ in (6).

 The separability condition (7) is explicitly solved, namely, 
the solution of the first inequality which satisfies the second 
constraint is given by
\begin{eqnarray}
c_{1}^{2}
&\leq&\frac{1}{4t^{2}}\{[2ab(1+t^{2})+t]-2\sqrt{D(a,b,t)}\},\nonumber\\
c_{2}^{2}
&\leq&\frac{1}{4}\{[2ab(1+t^{2})+t]-2\sqrt{D(a,b,t)}\}
\end{eqnarray}
with an auxiliary quantity  
\begin{eqnarray}
D(a,b,t)=a^{2}b^{2}(1-t^{2})^{2}+t(a+bt)(at+b)
\end{eqnarray}
for 
\begin{eqnarray}
0 \leq t\equiv |c_{2}|/|c_{1}|\leq 1
\end{eqnarray}
where we choose $|c_{2}|\leq |c_{1}|$ without loss of 
generality. When one defines 
\begin{eqnarray}
f(c^{2}_{1})=4(ab-c_{1}^{2})(ab-t^{2}c_{1}^{2})-(a^{2}+b^{2})-2tc^{2}_{1}
+\frac{1}{4}
\end{eqnarray}
for the first inequality in (7), $f(c^{2}_{1})$ assumes a negative minimum value
at $c_{1}^{2}=[2ab(1+t^{2})+t]/(4t^{2})>0$ and $f(0)=4(a^{2}-\frac{1}{4})(b^{2}-\frac{1}{4})\geq 0$. Thus $f(c^{2}_{1})=0$ has two non-negative solutions. Since the second constraint in (7) is written as $(2a-1)(2b-1)/(1+t)^{2}\geq c_{1}^{2}$, we examine
\begin{eqnarray}
&&f\left((2a-1)(2b-1)/(1+t)^{2}\right)\nonumber\\
&&=4x^{2}(a-\frac{1}{2})^{2}(b-\frac{1}{2})^{2}+4(a^{2}-\frac{1}{4})(b^{2}-\frac{1}{4})\nonumber\\
&&-4(a-\frac{1}{2})(b-\frac{1}{2})[ab(4-2x)+ \frac{1}{2}x]
\end{eqnarray}
with $0\leq x=4t/(t+1)^{2}\leq 1$, which is confirmed to be negative for 
 $0\leq x <1$ (for the generic case $a>\frac{1}{2}$ and $b>\frac{1}{2}$) and vanish for $x=1, i.e., t=1$. For $t=1$ the solution in (9) coincides with the second constraint in (7), and thus only the solution in (9) satisfies the second constraint in (7). Note that $|c_{1}|=|c_{2}|=0$ always defines
 a separable state.
 
Another essential ingredient in the proof of Simon~\cite{simon} 
is the P-representation for Gaussian systems. 
The Gaussian state is called P-representable if the density 
matrix is written as (we assume $\langle\hat{\xi}_{\mu}\rangle=0$ without loss
of generality)
\begin{eqnarray}
\hat{\rho}&=&\int d^{2}\alpha\int d^{2}\beta P(\alpha,\beta)
|\alpha,\beta\rangle\langle\alpha,\beta|
\end{eqnarray}
where $|\alpha,\beta\rangle$ is the coherent state defined by
$|\alpha,\beta\rangle=e^{\alpha\hat{a}^{\dagger}
-\frac{1}{2}|\alpha|^{2}}|0\rangle\otimes 
e^{\beta\hat{b}^{\dagger}-\frac{1}{2}|\beta|^{2}}|0\rangle$ with 
$\hat{a}=(\hat{q}_{1}+i\hat{p}_{1})/\sqrt{2}$ and $
\hat{b}=(\hat{q}_{2}+i\hat{p}_{2})/\sqrt{2}$.
Thus the P-representable states are separable.
By defining $\alpha=(\alpha_{1}+i\alpha_{2})/\sqrt{2}$ and 
$\beta=(\beta_{1}+i\beta_{2})/\sqrt{2}$, the weight factor 
$P(\alpha,\beta)$ in (14) is written in terms of the correlation 
matrix $V$ as
\begin{eqnarray}
&&P(\alpha,\beta)\\
&&=\frac{\sqrt{{\rm det} P}}{4\pi^{2}}
\exp\{-\frac{1}{2}(\alpha_{1},\alpha_{2},\beta_{1},\beta_{2})
P(\alpha_{1},\alpha_{2},\beta_{1},\beta_{2})^{T}\}\nonumber
\end{eqnarray}
with the matrix $P$  given by 
\begin{eqnarray}
P^{-1}=V-\frac{1}{2}I\geq 0
\end{eqnarray}
which defines the condition for the P-representation. See also
\cite{simon}.

The P-representation condition (16) is not invariant under 
$Sp(2,R)\otimes Sp(2,R)$ since $SS^{T}\neq I$ in general for
$S\in Sp(2,R)\otimes Sp(2,R)$. We thus examine the allowed range
of $Sp(2,R)\otimes Sp(2,R)$ transformations which preserve
the condition $S^{-1}V_{0}(S^{T})^{-1}-\frac{1}{2}I\geq 0$
or equivalently $V_{0}-\frac{1}{2}SS^{T}\geq 0$ by starting 
with the standard form in (3). We choose the 
$Sp(2,R)\otimes Sp(2,R)$ squeezing matrix $S$ as 
\begin{eqnarray}
S(r_{1},r_{2})S^{T}(r_{1},r_{2})=\left(\begin{array}{cccc}
  1/r_{1}&0&0&0\\
  0&r_{1}&0&0\\
  0&0&1/r_{2}&0\\
  0&0&0&r_{2}\\            
  \end{array}\right)
\end{eqnarray}
with suitably chosen $r_{1}\geq 1$ and $r_{2}\geq 1$.
The eigenvalues of 
$V_{0}-\frac{1}{2}S(r_{1},r_{2})S^{T}(r_{1},r_{2})$
are given by 
\begin{eqnarray}
(\lambda_{1})_{\pm}&=&\frac{1}{2}\{(a-\frac{1}{2r_{1}})+
(b-\frac{1}{2r_{2}})\nonumber\\
&&\pm
\sqrt{((a-\frac{1}{2r_{1}})-
(b-\frac{1}{2r_{2}}))^{2}+4c_{1}^{2}}\},\nonumber\\
(\lambda_{2})_{\pm}&=&\frac{1}{2}\{(a-\frac{1}{2}r_{1})+
(b-\frac{1}{2}r_{2})\nonumber\\
&&\pm\sqrt{((a-\frac{1}{2}r_{1})-(b-\frac{1}{2}r_{2}))^{2}
+4c_{2}^{2}}\}.
\end{eqnarray}
The P-representation exists if $(\lambda_{1})_{\pm}\geq 0$
and $(\lambda_{2})_{\pm}\geq 0$, namely, if the following
two conditions are simultaneously satisfied 
\begin{eqnarray}
&&(a-\frac{1}{2r_{1}})(b-\frac{1}{2r_{2}})\geq 
c_{1}^{2},
\nonumber\\
&&
(a-\frac{1}{2}r_{1})(b-\frac{1}{2}r_{2})\geq c_{2}^{2}
\end{eqnarray}
together with 
$(a-\frac{1}{2r_{1}})+(b-\frac{1}{2r_{2}})\geq 0$ and
$(a-\frac{1}{2}r_{1})+(b-\frac{1}{2}r_{2})\geq 0$. 

When one regards (9) and (19) as constraints on the 
pair of variables 
$(c_{1}^{2}, c_{2}^{2})$
for given $a$ and $b$,
the P-representation condition is more restrictive than 
the separability condition; since the P-representation
satisfies separablility, the set of points 
$(c_{1}^{2}, c_{2}^{2})$ allowed by the P-representation 
condition (19) always satisfy the separability condition 
(9). To be precise, we are working on the line defined by 
$t^{2}=c_{2}^{2}/c_{1}^{2}$. See also (30) below.
We thus expect  that these two conditions can coincide only for 
the extremal value of the P-representation condition (19) with 
respect to $r_{1}$ and $r_{2}$ with fixed $t$. 
 
We thus want to prove 
\begin{eqnarray}
&&(a-\frac{1}{2r_{1}})(b-\frac{1}{2r_{2}(t,r_{1})})\nonumber\\
&&=\frac{1}{t^{2}}
[(a-\frac{1}{2}r_{1})(b-\frac{1}{2}r_{2}(t,r_{1}))]\nonumber\\
&&=\frac{1}{4t^{2}}\{[2ab(1+t^{2})+t]-2\sqrt{D(a,b,t)}\}
\end{eqnarray}
for a suitable $1\leq r_{1}\leq 2a$ (and $1\leq r_{2}\leq 2b$)
for any given $0\leq t\leq 1$ by regarding $r_{2}$ as a 
function of $r_{1}$ and $t$. By this way we establish that the 
separability condition (9) agrees with the P-representation 
condition (19) with a suitable $Sp(2,R)\otimes Sp(2,R)$ 
transformation. 

We first consider the stationary points (or extremals) of
$(a-\frac{1}{2r_{1}})(b-\frac{1}{2r_{2}(t,r_{1})})$ and  
$(a-\frac{1}{2}r_{1})(b-\frac{1}{2}r_{2}(t,r_{1}))$ in (19)
with respect to 
$r_{1}$ with fixed $t$, namely
\begin{eqnarray}
&&(b-\frac{1}{2r_{2}(t,r_{1})})\frac{1}{r^{2}_{1}}
+(a-\frac{1}{2r_{1}})\frac{1}{r^{2}_{2}}\frac{\partial 
r_{2}}{\partial r_{1}}=0,\nonumber\\
&&(b-\frac{1}{2}r_{2}(t,r_{1}))+
(a-\frac{1}{2}r_{1})\frac{\partial r_{2}}
{\partial r_{1}}=0.
\end{eqnarray}
These two relations combined give rise to
\begin{eqnarray}
\frac{(ar_{1}-1/2)}{(a/r_{1}-1/2)}=
\frac{(br_{2}-1/2)}{(b/r_{2}-1/2)}.
\end{eqnarray}
The relation (22) combined with the first equality in (20)
 gives 
\begin{eqnarray}
&&r_{1}=\frac{\frac{r_{2}}{t}a+\frac{1}{2}}{a+\frac{1}{2}
\frac{r_{2}}{t}}, \ \ \
r_{2}=\frac{\frac{r_{1}}{t}b+\frac{1}{2}}{b+\frac{1}{2}
\frac{r_{1}}{t}}
\end{eqnarray}
which are symmetric in $r_{1}$ and $r_{2}$ and are solved as 
\begin{eqnarray}
r_{1}&=&\frac{1}{at+b}\{ab(1-t^{2})
+\sqrt{D(a,b,t)}\},\nonumber\\
r_{2}&=&\frac{1}{a+bt}\{ab(1-t^{2})
+\sqrt{D(a,b,t)}\}
\end{eqnarray}
with $0\leq t=|c_{2}|/|c_{1}|\leq 1$ and $D(a,b,t)$ defined in (10). The squeezing parameters 
are thus determined.

One can  confirm
\begin{eqnarray}
2a\geq r_{1}\geq 1, \ \ \ \ 2b\geq r_{2}\geq 1
\end{eqnarray}
for $a\geq \frac{1}{2}$ and $b\geq \frac{1}{2}$, if one combines the relations 
(23) with $\infty>r_{1}/t\geq 1$ and $\infty>r_{2}/t\geq 1$ valid for the solutions in (24). The relation $\infty>r_{1}/t\geq 1$, for example, is established if one
uses the expression 
\begin{eqnarray}
\frac{r_{1}}{t}=\frac{a+bt}{-ab(1-t^{2})+
\sqrt{a^{2}b^{2}(1-t^{2})^{2}+t(a+bt)(at+b)}}
\end{eqnarray}
together with
$t(at+b)\leq (a+bt)$  for $0\leq t \leq 1$ and the triangle inequality. 
Eq.(24) gives $r_{1}=r_{2}=1$ for $t=1$, and 
$r_{1}=2a, \ r_{2}=2b$ for $t=0$.

We emphasize that the condition (25) is required by a part of the P-representation condition $(a-\frac{1}{2r_{1}})+(b-\frac{1}{2r_{2}})\geq 0$ and
$(a-\frac{1}{2}r_{1})+(b-\frac{1}{2}r_{2})\geq 0$ in (19) 
when combined with (22) which implies $r_{2}(1)=1$ and 
$r_{2}(2a)=2b$ if one regards $r_{2}(r_{1})$ as a function of 
$r_{1}$. It should be noted that the squeezing
parameters themselves are essentially determined by the first equality in (20) which comes from the P-representation condition.

We finally evaluate by using $r_{1}$ and $r_{2}$ in (24) 
as
\begin{eqnarray}
&&\frac{1}{t^{2}}(a-\frac{1}{2}r_{1})(b-\frac{1}{2}r_{2})
\nonumber\\
&&=\frac{1}{4t^{2}}\{[2ab(1+t^{2})+t]-2\sqrt{D(a,b,t)}\}
\end{eqnarray}
which is a remarkable identity. This relation establishes 
(20), namely,
the fact that the boundaries of the conditions for separability 
and P-representation coincide for any 
$0\leq t=|c_{2}|/|c_{1}|\leq 1$.  It is significant that the squeezing parameters in (24) depend only on the ratio
$t=|c_{2}|/|c_{1}|$ and not on  $|c_{1}|$ and $|c_{2}|$ separately unlike the case in~\cite{duan}; this is because we are determining the bound to $|c_{1}|$ for given $t$ to be consistent with (9). All the states parameterized 
by $|c_{1}|$, which satisfy (9) for given $a,\ b \ {\rm and } \ t$, automatically satisfy the P-representation with the parameters in (24).

Our explicit construction proves that the 
P-representation condition (19) with the squeezing parameters in (24), which satisfy 
$1\leq r_{1}\leq 2a$ and $1\leq r_{2}\leq 2b$, is equivalent 
to the separability condition (9) for any 
$0\leq t=|c_{2}|/|c_{1}|\leq 1$, and thus the separability 
condition (9) is a necessary and sufficient separability
criterion for two-party Gaussian systems.
 To our knowledge,
our study is the first quantitative treatment of the basic
algebraic condition (7) which is considered to be 
fundamental~\cite{mancini2}. The existence of explicit analytic solutions of $r_{1}$ and $r_{2}$ in (24) for this basic problem is interesting, and the triplet 
\begin{eqnarray}
(V_{0},r_{1},r_{2})
\end{eqnarray}
 where  $V_{0}$ is the standard form in (3), characterize the general covariance matrix (2) of either separable or inseparable Gaussian states. 
These explicit 
solutions are convenient for quantitative treatment, and thus they should be 
useful for the quantitative  theoretical and experimental 
analyses of entanglement in practical applications.

\section{Implications of analytic solutions}

We would like to compare the present analysis with some of the past related works. 
The separability condition 
(4) is based on the partial transpose operation of the density 
matrix in the manner of Peres and Horodecki\cite{peres, horodecki}. 
The analysis of the negativity of the partial transposed density matrix has been further extended by Shchukin and Vogel~\cite{shchukin, miranowicz}. We here note a complementary property that 
the second moment for the separable density matrix
$\rho=\sum_{n}p_{n}\psi_{n}\psi^{\dagger}_{n}$ with $\psi_{n}=
\phi_{n}(q_{1})\varphi_{n}(q_{2})$ gives rise to a generalization of (6) 
\begin{eqnarray}
&&d^{T}Ad+f^{T}Bf+2d^{T}Cf+g^{T}Ag+h^{T}Bh+2g^{T}Ch\nonumber\\
&&\geq d^{T}\tilde{A}d+f^{T}\tilde{B}f+2d^{T}\tilde{C}f
+g^{T}\tilde{A}g+h^{T}\tilde{B}h+2g^{T}\tilde{C}h
\nonumber\\
&&+|d^{T}Jg| + |f^{T}Jh|
\end{eqnarray}
by an analysis similar to the derivation of the Kennard's 
uncertainty relation {\em without} referring to the partial transpose operation. Here $\tilde{V}$ in terms of $\tilde{A}$, $\tilde{B}$ and $\tilde{C}$ is defined analogously to (2) with the elements of the matrix 
$\tilde{V}=(\tilde{V}_{\mu\nu})$ defined by   
$\tilde{V}_{\mu\nu}=\sum_{k}p_{k}
\langle\Delta\hat{\xi}_{\mu}\rangle_{k}
\langle\Delta\hat{\xi}_{\nu}\rangle_{k}$ and $\Delta\hat{\xi}_{\mu}=\hat{\xi}_{\mu}-
\langle\hat{\xi}_{\mu}\rangle$. Note that one may choose $\langle\hat{\xi}_{\mu}\rangle=\sum_{k}p_{k}
\langle\hat{\xi}_{\mu}\rangle_{k}=0$, but the average $\langle\hat{\xi}_{\mu}\rangle_{k}$ for each component state  does not vanish in general.  
$\tilde{A}$ and $\tilde{B}$ are $2\times 2$ real symmetric
matrices and $\tilde{C}$ is a $2\times 2$ real matrix. Both of $V$ in (2) and $\tilde{V}$ are non-negative, and thus the condition (6) is necessary but not sufficient for separability in general. For the P-representation of the Gaussian state in (15) one can identify $P^{-1}=\tilde{V}$ by using a special property of the coherent state, which spoils $Sp(2,R)\otimes Sp(2,R)$ invariance due to normal ordering, and at the boundary of the P-representation condition where two eigenvalues of $P^{-1}$ vanish, the condition (29) can be equivalent to (6). \cite{footnote}.

One can also directly derive the condition (6) from the P-representation condition (16) which implies 
\begin{eqnarray}
d^{T}Ad+f^{T}Bf+2d^{T}Cf
-\frac{1}{2}(d^{T}d+f^{T}f)\geq 0
\end{eqnarray}
for any $d$ and $f$. When one adds (30) to a relation obtained from (30)
by replacing $d$ and $f$ by $g$ and $h$, respectively, 
one recovers the separability condition (6)  
\begin{eqnarray}
&&d^{T}Ad+f^{T}Bf+2d^{T}Cf+g^{T}Ag+h^{T}Bh+2g^{T}Ch\nonumber\\
&&\geq
\frac{1}{2}(d^{T}d+f^{T}f)+\frac{1}{2} (g^{T}g+h^{T}h)
\geq |d^{T}Jg|+|f^{T}Jh|,\nonumber
\end{eqnarray}
where we used the relation such as 
$\frac{1}{2}(d^{T}d+g^{T}g)
\geq \sqrt{(d^{T}d) (g^{T}J^{T}Jg)}
\geq |d^{T}Jg|$.

What we have proved in the present paper (and also in \cite{simon}) is that (6) also implies (16) with the help of the squeezing operation.
In this proof, the second weaker condition in (7) played a crucial role, namely, the first relation in (7) does not encode the entire information of the separability condition (4) or (6).
The importance of the second condition in (7), which appears to be overlooked in the past related works (see, for example, eq.(19) in \cite{simon}, Theorem V.2 in \cite{mancini2} and eq.(23) in 
\cite{shchukin}), was first recognized clearly in the present explicit analytic treatment of the separability condition (6).

As for the second weaker condition in (7), which is not $Sp(2,R)\otimes Sp(2,R)$ invariant, one can confirm that  this weaker condition 
\begin{eqnarray}
\sqrt{[ar_{1}+\frac{a}{r_{1}}-1][br_{2}+\frac{b}{r_{2}}-1]}
\geq \sqrt{r_{1}r_{2}}|c_{1}|+\frac{|c_{2}|}{\sqrt{r_{1}r_{2}}}
\end{eqnarray}
written for the second moment, which is obtained from the standard form $V_{0}$ in (3) by a squeezing operation $S^{-1}$  in (17), corresponds to the separability condition in \cite{duan}; in fact, if one imposes the condition (22),
 the left-hand side of (31) becomes 
\begin{eqnarray}
&&\sqrt{[ar_{1}+\frac{a}{r_{1}}-1][br_{2}+\frac{b}{r_{2}}-1]}
\\
&=&\sqrt{(ar_{1}-1/2)(br_{2}-1/2)} + \sqrt{(a/r_{1}-1/2)(b/r_{2}-1/2)}\nonumber
\end{eqnarray}
and one recovers eq. (16) in \cite{duan} when converted into their notation. 

From our analysis of (13),
it is obvious that some parameter region allowed by the weaker 
condition (31) with $r_{1}=r_{2}=1$ does not satisfy the P-representation condition. But if one combines this weaker relation with suitable squeezing, a simple direct proof of the P-representation is obtained. If one sets $|c_{2}|=t|c_{1}|$ in (31) with the condition (22), one obtains
\begin{eqnarray}
&&\sqrt{(ar_{1}-1/2)(br_{2}-1/2)} + \sqrt{(a/r_{1}-1/2)(b/r_{2}-1/2)}\nonumber\\
&&\geq [\sqrt{r_{1}r_{2}}+\frac{t}{\sqrt{r_{1}r_{2}}}]|c_{1}|.
\end{eqnarray}
If one uses the analytic formulas of squeezing parameters given in (24), one can confirm that the relation (33) when regarded as a bound to $|c_{1}|$
becomes identical to the $Sp(2,R)\otimes Sp(2,R)$ invariant bound (9) and also to the boundary of the P-representation condition (19). Here we use (20) and its square root, namely,
\begin{eqnarray}
&&\frac{\sqrt{(ar_{1}-1/2)(br_{2}-1/2)}}{\sqrt{r_{1}r_{2}}}\nonumber\\
&&=\frac{ \sqrt{(a/r_{1}-1/2)(b/r_{2}-1/2)}}{(t/\sqrt{r_{1}r_{2}})}
\nonumber\\
&&=\frac{1}{2t}\{[2ab(1+t^{2})+t]-2\sqrt{D(a,b,t)}\}^{1/2}.
\end{eqnarray}
Since (19) implies (33), one concludes that the weaker separability condition combined with our explicit squeezing parameters provides the necessary and sufficient condition for the P-representation. In a symbolic notation we have
\begin{eqnarray}
{\rm eq.}(33) \supseteq {\rm eq.}(9) \supseteq {\rm eq.}(19),
\end{eqnarray}
which means, for example, any standard form of the covariance matrix $V_{0}$ which satisfies (19) automatically satisfies (9) (a stronger condition means a smaller set of $V_{0}$). But our analysis above shows that (33) and (19), both of which depend on squeezing parameters,
coincide with (9) for our explicit solutions of squeezing parameters.
This illustrates the power of our explicit analytic formulas, and this simple proof of the P-representation gives  another explicit proof of the necessary and sufficient condition for separability of two-party Gaussian systems on the basis of (33).
 
 We next briefly discuss the quantitative difference between our scheme and the scheme in \cite{duan}. The authors in \cite{duan}
look for the solution $f(r^{\star}_{1})=0$ where 
\begin{eqnarray}
f(r_{1})&=&
[\sqrt{r_{1}r_{2}}|c_{1}| - \sqrt{(ar_{1}-1/2)(br_{2}-1/2)}]
\\
&-&[|c_{2}|/\sqrt{r_{1}r_{2}} - 
\sqrt{(a/r_{1}-1/2)(b/r_{2}-1/2)}]\nonumber
\end{eqnarray}
and the condition (22) when written in our notation. Namely,
they look for the solution of 
\begin{eqnarray}
&&\sqrt{(ar_{1}-1/2)(br_{2}-1/2)} - \sqrt{r_{1}r_{2}}|c_{1}|
\nonumber\\
&&=\sqrt{(a/r_{1}-1/2)(b/r_{2}-1/2)} - |c_{2}|/\sqrt{r_{1}r_{2}}
\end{eqnarray}
together with (22) and (25), though the condition (25) is not explicitly stated in \cite{duan}. In contrast, in our scheme we solve 
\begin{eqnarray}
&&\frac{\sqrt{(ar_{1}-1/2)(br_{2}-1/2)}}{\sqrt{r_{1}r_{2}}|c_{1}|}\nonumber\\
&&=\frac{\sqrt{(a/r_{1}-1/2)(b/r_{2}-1/2)}}{(|c_{2}|/\sqrt{r_{1}r_{2}})}
\end{eqnarray}
together with (22) and (25), as is seen from the first equality of (34).
Since (38) implies 
\begin{eqnarray}
&&\frac{\sqrt{(ar_{1}-1/2)(br_{2}-1/2)}-\sqrt{r_{1}r_{2}}|c_{1}|}{\sqrt{r_{1}r_{2}}|c_{1}|}\nonumber\\
&&=\frac{\sqrt{(a/r_{1}-1/2)(b/r_{2}-1/2)}-|c_{2}|/\sqrt{r_{1}r_{2}}}{(|c_{2}|/\sqrt{r_{1}r_{2}})},
\end{eqnarray}
the {\em common} solutions of (37) and (38) exist only for 
\begin{eqnarray}
&&\sqrt{(ar_{1}-1/2)(br_{2}-1/2)} - \sqrt{r_{1}r_{2}}|c_{1}|
\\
&&=\sqrt{(a/r_{1}-1/2)(b/r_{2}-1/2)} - |c_{2}|/\sqrt{r_{1}r_{2}}
=0\nonumber
\end{eqnarray}
or else for $\sqrt{r_{1}r_{2}}|c_{1}|=|c_{2}|/\sqrt{r_{1}r_{2}}$,
namely,
\begin{eqnarray}
r_{1}r_{2}=\frac{|c_{2}|}{|c_{1}|}=t.
\end{eqnarray}
The relation (40) together with our explicit analytic solutions of $r_{1}$ and $r_{2}$ shows that the values of $|c_{1}|$ and $|c_{2}|$ given by (40) correspond to the largest allowed values of $|c_{1}|$ and $|c_{2}|$ in the separability condition (9) if one recalls (34). As for (41), we can choose 
$0\leq t\leq  1$ without loss of generality as in (11) and the squeezing parameters are bounded as in (25) by the P-representation condition when combined with (22). Thus the condition (41) is satisfied only at $t=1$ and $r_{1}=r_{2}=1$.
The two conditions (37) and (38) thus coincide only for such a measure-zero subset of the standard form of separable covariance matrices $\{V_{0}\}$. This quantitative difference between the two schemes is interesting.  

As we explained already, the squeezing parameters in our scheme 
specify the boundary of the P-representation condition for any given $t$, which agrees with  the separability condition (9). On the other hand, the scheme in \cite{duan} specifies the squeezing parameters for each given covariance matrix separately such that the P-representation condition is ensured, and thus the specification of squeezing parameters is more local. Our analytic solutions depend only on the ratio $t=|c_{2}|/|c_{1}|$, while squeezing parameters in (37) generally depend on $|c_{1}|$ and $|c_{2}|$ separately. Nevertheless, we here show that our explicit analytic solutions allow a concrete proof of the separability criterion in the scheme of \cite{duan}. For this purpose, we extend (35) to a symbolic notation 
\begin{eqnarray}
{\rm eq.}(33) \supseteq {\rm eq.}(9) \supseteq {\rm eq.}(19)
\supseteq \{{\rm eq.}(33) \cap {\rm eq.}(37)\}
\end{eqnarray} 
where the last relation means that any $V_{0}$ which satisfies 
(37) in addition to (33) automatically satisfies the P-representation condition (19), as is explicitly confirmed \cite{duan}.
Thus $\{{\rm eq.}(33) \cap {\rm eq.}(37)\}$ provides a {\em sufficient} condition for the P-representation, but the converse
is not obvious. The P-representation condition (19) implies (33)
as is shown in \cite{duan}, but it is not obvious if (19) implies (37). To be more precise, it is not obvious if (9) implies (37). In fact, the authors in \cite{duan} choose a rather general class of $V_{0}$ and prove the solution of $f(r^{\star}_{1})=0$ in the domain $1\leq r^{\star}_{1}<\infty$
which, however, does not satisfy a part of the P-representation condition (25), namely, $1\leq r^{\star}_{1}\leq 2a$.

We now show that (9) implies (37) by using our analytic solutions. We first write (36) in the form  
\begin{eqnarray}
&&f(r_{1}, |c_{1}|)=
(\sqrt{r_{1}r_{2}}-t/\sqrt{r_{1}r_{2}})|c_{1}| \\
&&- \sqrt{(ar_{1}-1/2)(br_{2}-1/2)}
+ \sqrt{(a/r_{1}-1/2)(b/r_{2}-1/2)}\nonumber
\end{eqnarray}
together with (22). We then have $f(r_{1}=1, |c_{1}|)>0$ for $0\leq t<1$ (the case $t=1$ is trivial since then $r_{1}=r_{2}=1$ is the solution). From the analysis of (40), we have $f(r_{1}, |c_{1}|)=0$ for the largest allowed $|c_{1}|$ in (9) and for our analytic solutions. Since $\sqrt{r_{1}r_{2}}\geq 1$, the function $f(r_{1}, |c_{1}|)$ is a linear increasing function of $|c_{1}|$ when we fix squeezing parameters at our analytic solutions. Here it is crucial that our analytic solutions (24) depend only on $a,\ b$ and $t$. We thus conclude  
\begin{eqnarray}
f(r_{1}, |c_{1}|)\leq 0
\end{eqnarray}
for all values of $|c_{1}|$ in (9) when the squeezing parameters 
are fixed at our analytic solutions.
Since our analytic solutions satisfy $1\leq r_{1} \leq 2a$ as in (25), the relation (44) combined with $f(r_{1}=1, |c_{1}|)>0$ shows that $f(r^{\star}_{1})=0$ has a solution in the interval $1\leq r^{\star}_{1} \leq 2a$ for all values of $|c_{1}|$ in (9). Namely, (9) implies (37). This completes the proof of the necessary and sufficient separability condition in the scheme of \cite{duan}.

In the above analysis, we used the solution of (22)
\begin{eqnarray}
r_{2}(r_{1})=\frac{4b}{[\sqrt{(1-X)^{2}+16b^{2}X}+(1-X)]}
\end{eqnarray}
with $X(r_{1})=(2a/r_{1}-1)/(2ar_{1}-1)$
which assumes $X(1)=1$ and $X(2a)=0$, and 
thus $r_{2}(1)=1$ and $r_{2}(2a)=2b$.

\section{Discussion}

We found the explicit analytic formulas of squeezing parameters which establish the equivalence of the $Sp(2,R)\otimes Sp(2,R)$ invariant separability condition with the P-representation condition. These explicit analytic formulas give rise to not
only the concrete proofs of the separability criterions for two-party Gaussian systems formulated in \cite{simon} and \cite{duan} but also a new simple proof of the separability criterion 
as is described in connection with (35). Our analytic formulas thus provide a unified basis for the analysis of separability in continuous variable two-party Gaussian systems.
    
We analyzed the two-party system with one freedom for each party. The system we analyzed may be more properly called a two-mode system since the analysis of the two-party system with more than one freedom in each party is more involved \cite{werner}. As for the non-negativity of the partially transposed 
density matrix, 
Vidal and Werner \cite{vidal} analyzed the separability of Gaussian states by using logarithmic negativity, which is essentially the same as the actual analysis of Simon \cite{simon} and leads to (7) and (16),
but no analytic formulas of squeezing parameters are given. See also the related references \cite{englert} -
\cite{giedke1}.

\bibliography{apssamp}

\begin{thebibliography}{99}
\bibitem{footnote}
A  more detailed account of this issue will be given elesewhere.
\end{thebibliography}

\end{document}